\documentclass[a4paper]{jpconf}
\usepackage{graphicx}
\usepackage{floatrow}
\DeclareFloatSeparators{mysep}{\hspace{1cm}}

\begin{document}
\title{The time as an emergent property of quantum mechanics, a synthetic description of a first experimental approach}
\author{E Moreva$^{1,2}$, G Brida$^1$, M Gramegna$^1$, V Giovannetti$^3$, L Maccone$^4$ and  M Genovese$^{1*}$}
\address{$^1$INRIM, strada delle Cacce 91, 10135 Torino, Italy}
\address{$^2$International Laser Center of M.V.Lomonosov Moscow State University, 119991, Moscow, Russia}
\address{$^3$NEST, Scuola Normale Superiore and Istituto
    Nanoscienze-CNR, piazza dei Cavalieri 7, I-56126 Pisa, Italy}
\address{$^4$Dip.~Fisica ``A.~Volta'', INFN Sez.~Pavia, Univ.~of Pavia,
  via Bassi 6, I-27100 Pavia, Italy}
\ead{$^{*}$m.genovese@inrim.it}

\begin{abstract}
The "problem of time" in present physics substantially consists in the fact that a straightforward quantization of the general relativistic evolution equation and constraints generates for the  Universe wave function the Wheeler-De Witt equation, which describes a static Universe. Page and Wootters  considered the fact that there exist states of a system composed by entangled subsystems that are stationary, but one can interpret the component subsystems as evolving: this leads them to suppose that the global state of the universe can be envisaged as one of this static entangled state, whereas the state of the subsystems can evolve.
Here we synthetically  present  an experiment, based on PDC polarization entangled photons, that shows a practical example where this idea works, i.e. a subsystem of an entangled state works  as a "clock" of another subsystem.

\end{abstract}

\section{Introduction}

 {\em \bf ``...an infinite series of times, in a dizzily growing, ever
  spreading network of diverging, converging and parallel times. This
  web of time--the strands of which approach one another, bifurcate,
  intersect or ignore each other through the centuries--embraces
  every possibility.'' [Borges]}

  {\em ``Quid est ergo tempus? si nemo ex me quaerat, scio;    si quaerenti explicare velim, nescio.''} \cite{a}
As Augustinus Hipponensis present physicists are in the situation where time is an essential physical parameter whose meaning is intuitively clear, but several problems arise  when they try to provide a clear definition of time. First of all the definition of time is different in different
branches of physics as classical and non-relativistic quantum mechanics (a  fixed background parameter), special relativity (a proper time for each observer, time as fourth coordinate; set of privileged inertial frames) or general relativity
(time is a general spacetime coordinate: not at all an absolute time, time non orientability, closed timelike curves)\cite{Anderson}.
More specifically, the problem of time \cite{Anderson, Sorkin, kuchar, isham, ashtekar, altra, nik, nic, alt, y} in physics consists in the fact that a straightforward quantization of the general relativistic evolution equation and constraints generates the Wheeler-De Witt equation $H_{tot}\left|\Psi\right\rangle = 0$, where $H_{tot}$ is the global Hamiltonian of the universe and $\left|\Psi\right\rangle $ is its state.
Obviously, this means that the state of the universe must be static, which clashes with our everyday experience of an evolving world.

Apart from the Wheeler-De Witt equation, one can also consider that a time shift of the state of the whole universe must be unobservable from a purely physical consideration: if shifting the state of the whole universe, there is nothing external that can keep track of this shift, so this shift must be unobservable (it is a trivial application of Mach's
principle). Same considerations would apply, of course, also to the spatial degrees of freedom: a global shift in position of the whole state of the universe must similarly be unobservable, so the universe must be in an eigenstate of its momentum operator. In the following we will only consider temporal degrees of freedom.

Incidentally, this idea of a static universe surprisingly re-proposes, of course rephrased in modern language, ideas stemming from the work of Parmenides of Elea ["the phenomena of movement and change are simply appearances of a static, eternal reality"] and then diffused in roman-hellenistic word ["Tempus item per se non est, sed rebus ab ipsis consequitur sensus, transactum quid sit in aevo, tum quae res instet, quid porro deinde sequatur" \footnote{Time also exists not by itself, but simply from the things which happen the sense apprehends what has been done in time past, as well as what is present and what is to follow after} Titus Lucretius Carus, De Rerum Natura].

The Page and Wootters scheme \cite{Wootters, Page}, which developed some previous ideas \cite{previous}, is based on the fact that there exist states of a system composed by entangled subsystems that are stationary, but one can interpret the component subsystems as evolving.
One can then suppose that the global state of the universe as one of this static entangled state, whereas the state of the subsystems (us, for example) can evolve. This solves in an extremely elegant way the problem of time.
Incidentally, Page and Wootters' proposal naturally embodies the philosophy of relationalism \cite{rovelli,rovt} and
operationalism, since time is only defined in relation to clocks and to its measurement procedure \cite{rudolphrmp,wiseman}.

Up to now, all these considerations were of theoretical character. Here we epitomize an experimental approach \cite{pra} to this problem by providing an emblematic example
of  Page and Wootters idea at work, visualizing how time could emerge
from a static (with respect to an abstract external time) entangled state.
Even though the total state of a system is static, time is recovered as correlations between a subsystem that acts as a clock and the rest of the system that
evolves according to such a clock. We use a system composed of two entangled photons (a paradigmatic system for several emblematic experiments \cite{2f}): the rotation of the polarization of the first acts as a clock for proving
an evolution of the polarization of the second. Nonetheless, we demonstrate that the joint polarization state of both photons does not evolve.
The Page-Wooters mechanism has been criticised \cite{kuchar,unruh}, and Gambini et al. have proposed some new ideas \cite{gambinipullin,decoherence} to overcome these criticisms. Out experiment demonstrates also some aspects of these ideas.

\section{The experiment}

The experimental setup is schematically depicted in Figure 1. It consists of two blocks, "preparation" and "measurement".
The preparation block allows producing a family of ququarts biphoton polarization states \cite{m} of the form:

\begin{equation}
\left|\Psi\right\rangle=\cos\theta\left|{HH}\right\rangle + e^{i\varphi} \sin\theta \left|{VV}\right\rangle
\label{eq:state}
\end{equation}

\begin{figure}[ht] \begin{center}
\includegraphics[width=0.6\textwidth]{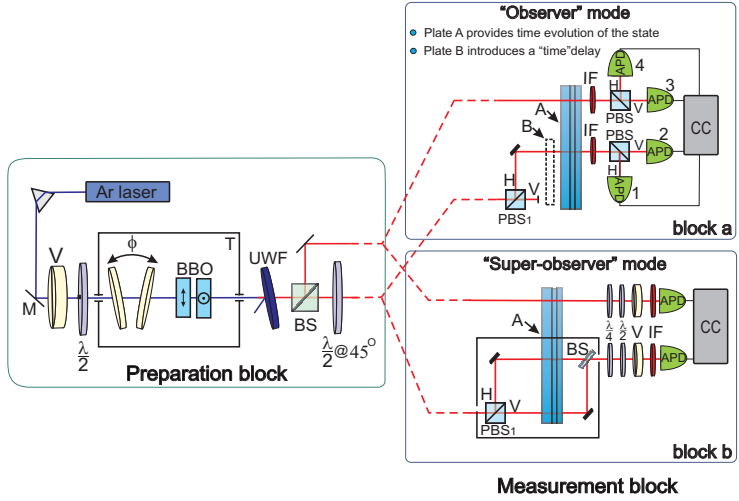}
\caption{Scheme for the two experiments. The experimenter in observer mode (block a) can prove the
  time evolution of the first photon (upper path) using only
  correlation measurements between it and the clock photon (lower
  path) without access to an external clock. The experimenter in
  super-observer mode (block b) proves that the global state of the
  system is static through state tomography. The blue boxes (A) represent different thicknesses of
  birefringent plates which evolve the photons by rotating their
  polarization: different thicknesses represent different time
  evolutions. The dashed box (B) represents a (known) phase delay of the clock photon only; PBS
  stands for polarizing beam splitter in the $H/V$ basis; BS for beam splitter.}\label{f:schema}
\end{center}
\end{figure}

It includes two orthogonally oriented BBO crystals ($1 mm$) that, pumped by a $700$ mW cw Ar laser operating at $351$ nm, generate a pair of the basic ququart states via type-I spontaneous parametric down conversion in collinear, frequency degenerate regime around the central wavelength of $702$ nm. The basic state amplitudes are controlled with the help of a Thompson prism (V), oriented verticaly and the half-wave plate $\lambda/2$ at an angle $\theta$.
Two $1$ mm quartz plates (QP), that can be rotated along the optical axis, introduce a phase shift $\varphi$ between horizontally and vertically polarized type-I entangled \cite{gio} biphotons.
Non-linear crystals and quartz plates are placed into a temperature-stabilized closed box for achieving stable phase-matching conditions all the time. The preparation block also includes a non-polarizing beam splitter (BS), which is used to split the initial (collinear) biphoton field into two spatial modes $1,2$.
For preparing the singlet Bell state

\begin{equation}
\left|{\Psi^-}\right\rangle=\frac{1}{{\sqrt2}}\left({\left|H_{1}\right\rangle\left|V_{2}\right\rangle-\left|V_{1}\right\rangle\left|H_{2}\right\rangle}\right)
\label{eq:singlet}
\end{equation}

parameters $\theta=45^\circ, \varphi=0^\circ$, were selected and an additional half-wave plate $\lambda/2$ at $45^\circ$ was introduced in the transmitted arm.

The measurement part can be mounted in two different modes "observer" and "super-observer". "Observer" mode (Figure 1, block a) serves as proving that one subsystem (polarization of the upper photon) evolves with respect to a clock constituted by the other subsystem (polarization of the lower photon).
In the "super-observer" mode ((Figure 1, block b)) we show that the global state of the two photons is static with respect to any "external" clock. Let us analyze the observer mode first.

\textbf{"Observer" mode:} Each arm of a measurement block contains interference filters (IF) with central wavelength $702$ nm (FWHM $1$ nm) and polarizing beam splitter (PBS). Four avalanche photodiode detectors (APD) are placed at the ends and connected to a coincidence count scheme (CC) with $1.5$ ns time window.

In this mode the polarization of the photon in the lower arm is used as a clock for proving an evolution of the polarization of the photon in the upper arm.
In other words, the first polarizing beam splitter PBS acts as a non-demolition measurement that measures (in the H/V basis) the polarization of the second photon in the state $\left|{\Psi^-}\right\rangle$ (\ref{eq:singlet}). Once it is initialized in the $H$ state, the polarization of both photons evolves in the birefringent quarts plates (A) at $45^\circ$  as $\left| H \right\rangle \to \frac{1}{{\sqrt 2 }}\left( {\left| H \right\rangle + e^{i\delta } \left| V \right\rangle } \right),\left| V \right\rangle \to \frac{1}{{\sqrt 2 }}\left( {\left| H \right\rangle  - e^{i\delta } \left| V \right\rangle } \right)$, where $\delta$
 is the materials optical thickness. The clock readout is performed on the lower detectors, which measure whether the photon is $\left| H \right\rangle$ or $\left|V\right\rangle$.
This clock is extremely primitive, since it has a dial with only two values: either it found with $H$ polarization, corresponding to time $t_1$ or with $V$ polarization, corresponding to time $t_2$.

In the "observer" mode we do not have access to any external clock, the only information we obtain are correlations (coincidences) between the detectors. Thus, this primitive clock allows extracting the information about the polarization of the upper photon in a defined state, for example $\left|V\right\rangle$, only from registering coincidence counts between the detectors $1-3$ (that corresponds to the measurement at the moment time $t_1$ ) or 2-3 (that corresponds to time $t_2$). Furthermore, one does not have any knowledge about the past time after the clock initialization (knowledge about this time is equivalent to the presence of an external clock), so we have to record all events taking place in selected points of time $t_1$ and $t_2$. From the experimental point of view it means measurements of coincidences between the detectors 1-3 and 2-3 for  all possible thicknesses of the birefringent plates A \cite{gambinipullin,montev,pagereply}. Averaging the number of counts from detectors 1-3 and 2-3 over the total number of photon pairs from all detectors 1-2-3-4, gives the probability of finding the upper photon in polarization state $\left|V\right\rangle$ $p(t_1)$ and $p(t_2)$ at two points of time.

\begin{figure}[!ht]
\TopFloatBoxes
\begin{floatrow}

\ffigbox[\FBwidth]{\caption{Experimental results. Probability $p(t)$ that the upper photon is $V$ as a function of the time $t$ recovered from the lower photon. The points with matching colors represent $p(t_1+\tau_i)$ and $p(t_2+\tau_i)$: yellow, red, blue, etc., for $i=0,1,2,\cdots$, respectively. Here nine different values of $\tau_i$ are obtained from a thick quartz plate rotated by nine different angles. The dashed line is the theoretical value.}}{\includegraphics[width=0.55 \textwidth]{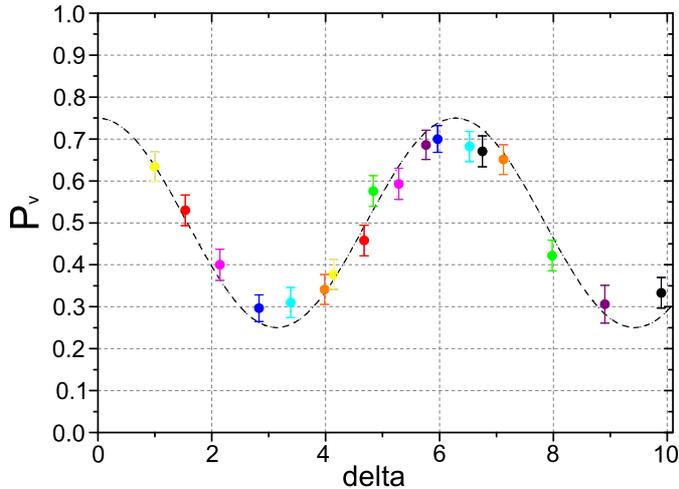}}\label{f:resultsgp}
\killfloatstyle

\ttabbox[\Xhsize]{\caption{Bell state reconstruction. The table contains values of "external" time and fidelities.}}
{\begin{tabular}{ll}

\br
"external" time & Fidelity \\
\mr
0 & 0.953 \\
\mr
$\pi/2\omega$ & 0.928 \\
\mr
$\pi/4\omega$ & 0.940  \\
\mr
$3\pi/2\omega$ & 0.932 \\
\br
\end{tabular}
}
\end{floatrow}
\end{figure}

To obtain a more interesting clock, we perform the same conditional probability measurement introducing varying time delays to the clock photon, implemented through quartz plates of variable thickness (dashed box B in Figure ~\ref{f:schema}, block b).  [Even though he has no access to abstract coordinate time, he can have access to systems that implement known time delays, that he can calibrate separately.]  Now, we obtain a sequence of time-dependent values for the conditional probability:
$p(t_1+\tau_i)=P^{\tau_i}_{3|1}$ and $p(t_2+\tau_i)=P^{\tau_i}_{3|2}$, where $\tau_i=\delta_i/\omega$ is the time delay of the clock photon obtained by inserting the quartz plate B with thickness $\delta_i$ in the clock photon path. The experimental results are presented in Fig.~\ref{f:resultsgp}, where each colour represents a different delay: the yellow points refer to $\tau_0$; the red points to $\tau_1$, etc.


\thisfloatsetup{floatrowsep=mysep}
\emph{}

Our results are in very good agreement with the theory (dashed line) \cite{pra}. The reduction in visibility of the sinusoidal time dependence of the probability is caused by the decoherence effect due to the use of a low-resolution clock (our clock outputs only two possible values), a well known effect \cite{decoherence,montev,cld}.

\textbf{"Super-observer" mode:} The former parter of the experiment is addressed to show that one photon "evolves" when using the other photon as a clock.
Now we have to show that the global state of two photons is static with respect to any "external" clock. To do so, we need to repeat the whole experiment in "super-observer" mode, and show that for arbitrary thicknesses of the birefringent plates which represent an "external" time the state is invariant. This setup of the experiment is shown in Figure 1(block b). Here the  $50/50$ beam splitter (BS) on the output of a balanced interferometer is performing a quantum erasure of the polarization measurement performed by the polarizing beam splitter PBS1. In the ideal case the beam splitter should coherently transmit a photon that passes in the upper arm and reflects a photon that passes in the lower arm of the interferometer without affecting its polarization. Without loss of generality, in our setup we use a conventional beam splitter and post-selection on one of the exits. For temporal stability the interferometer was placed into a closed box and narrow interference filters ($702$ nm, FWHM=$1$ nm) were inserted before the detectors. A set of quartz plates (A) with different thicknesses are used to introduce an "external" time.

For analyzing an output state of transformed ququarts we used the quantum state tomography procedure. Necessary projective measurements were realized by polarization filters placed in front of detectors. Each filter consists of a sequence of quarter- and half-wave plates and a polarization prism which transmits vertical polarization. Two APDs linked to a coincidence scheme with 1.5 ns time window were used as single photon detectors.
In our experiment we used the protocol of Ref. \cite{kwiat2001,our}. Registering the coincidence rate for 16 different projections, that were realized by half and quarter plates and a fixed analyzer, it was possible to reconstruct the arbitrary polarization state of ququarts.

The accuracy of quantum tomography is estimated through the fidelity $F = \left({Tr\sqrt {\rho _{in}^{1/2} \rho _{out} \rho _{in}^{1/2} } }\right)^2$ where $\rho_{in}$ is the initial (theoretical) density matrix and $\rho_{out}$ is the reconstructed density matrix after transformation.
Our results are summarized in Table I, demonstrating the stationarity of the global state respect to the evolution.


\section{Conclusions}
In summary, by running our experiment in two different modes (``observer'' and ``super-observer'' mode) we have experimentally
shown how the same energy-entangled Hamiltonian eigenstate can be perceived as evolving by the internal observers that test the
correlations between a clock subsystem and the rest (also when considering two-time measurements), whereas it is static for the
super-observer that tests its global properties. Our experiment is a practical implementation of the Page and Wooters\cite{pw} and Gambini et al. \cite{gambinipullin} mechanisms but,
obviously, it cannot discriminate between this and other proposed solutions for the problem of time \cite{kuchar,isham,ashtekar,Anderson,altra}, representing only an "illustration" of this phenomenon. Developments to higher dimensional systems could eventually be envisaged by using atoms \cite{mas}.

In closing, we note that the time-dependent graphs of Fig.~\ref{f:resultsgp} have been obtained without any reference to an external time (or phase) reference, but only from measurements of correlations between the clock photon and
the rest representing an implementation of a `relational' measurement of a physical quantity (time) relative to an internal quantum reference frame \cite{rudolphrmp,wiseman}, i.e. an example of relational metrology.

\section*{Acknowledgments}
This research was supported by  the John Templeton Foundation (the opinions expressed in this publication are those of the
authors and do not necessarily reflect the views of the John Templeton Foundation), E.V.Moreva acknowledges the support from the Dynasty Foundation and Russian Foundation for Basic Research (project 13-02-01170-a).

\end{document}